\documentclass[prl,aps,showpacs,twocolumn]{revtex4}%
\usepackage{graphicx}
\usepackage{amsmath}
\usepackage{amsfonts}
\usepackage{amssymb}%
\providecommand{\U}[1]{\protect\rule{.1in}{.1in}}
\providecommand{\U}[1]{\protect \rule{.1in}{.1in}}
\providecommand{\U}[1]{\protect \rule{.1in}{.1in}}
\providecommand{\U}[1]{\protect \rule{.1in}{.1in}}
\begin{document}
\title{Magnetic Merry Go Round - Resonant Reshaping of Colloidal Clusters on a Current Carrying Wire}
\author{Lydiane Becu$^{1}$}
\author{Marc Basler$^{2}$}
\author{Miodrag L. Kuli\'{c}$^{3,4}$}
\author{Igor M. Kuli\'{c}$^{2}$}
\email{kulic@unistra.fr}
\date{\today}

\begin{abstract}
We describe a simple physical method for trapping and transforming
magnetic colloid clusters on a current carrying wire. We use the
wire's field as a mould to form colloidal rings and helices on its
surface. To transform the initially amorphous, kinetically
trapped, bulky clusters we induce a low frequency magnetic
modulation wave that spins around the wire axis, effectively
eliminates defects from the clusters and stretches them into
slender rings and helical structures. A qualitative theoretical
model of the underlying resonant transformations is developed and
the practical potential of the wire as a magnetic micro-assembler
is discussed.

\end{abstract}
\affiliation{$^{1}$Universit{\'e} de Lorraine, LCP-A2MC, Institut de Chimie, Physique et
Mat{\'e}riaux, 1 Bd. Arago, 57070 Metz, France}
\affiliation{$^{2}$CNRS, Institute Charles Sadron, 23 rue du Loess BP 84047, 67034
Strasbourg, France }
\affiliation{$^{3}$ Institute for Theoretical Physics, Goethe-University D-60438 Frankfurt
am Main, Germany}
\affiliation{$^{4}$Institute of Physics, Belgrade, Serbia}

\maketitle

\section{Introduction}

Assembling controlled super-structures on the micro- and nanoscale
is among the current challenges in colloidal science. Many
original methods to guide the formation of well-defined colloidal
structures have been invented, including assembly of colloidal
rings at interfaces \cite{RingsInMixture} and in ferrofluids
\cite{RingsInFerrofluids}, colloidal helices via chiral templating
\cite{BibetteChiralCLusters} and complex Janus particle
interactions \cite{RingsInJanusRods}. Effective interactions
between magnetic beads via dynamic magic angle spinning
\cite{Martin}, giving rise to novel many body, Van der Waals-like
interactions \cite{Kulic1} have been utilized in forming
self-healing membranes and foams \cite{Osterman} that are finding
first applications in templating cellular tissue growth
\cite{TISSUESlov}.

In contrast to the equilibrium, thermally driven self-assembly,
the process of a non-equilibrium field driven assembly induces
interactions that are often stronger than the thermal energy scale
\cite{TiernoReview}. If the energy landscape is sufficiently
simple and smooth, large field driven, effective interactions can
speed up and direct the kinetics and furthermore open pathways to
new dissipative structures, and non-equilibrium steady states
\cite{fda_Review,Snezhko}. However, in most cases where the energy
landscape is rugged and has many meta-stable states, large
barriers effectively prevent the system from finding a unique
configuration in experimentally practical times. Lacking the
simplifying principle of free energy minimization one is forced to
resort to new ideas for guiding the colloidal assemblies to their
desired final structure.

In this paper we expand the repertoire of ideas for tailoring colloidal
clusters by a new dynamical mechanism. We force randomly formed clusters into
desired shapes - in our case colloidal rings- by applying out-off equilibrium
forces via dynamic fields. Linear colloidal super-structures like rings and
helices have an esthetic appeal. Yet, owing to their non-straight geometry,
they appear notoriously difficult to generate. Despite minimizing magnetic
energy, magnetic rings are rarely observed spontaneously
\cite{MagnetotacticBact} and require spacial conditions, like confinement on
solid surfaces \cite{RingsInFerrofluids} and liquid interfaces
\cite{RingsInMixture} to form. One interesting idea that we stumbled upon and
present in this paper is to use a "field moulding" method on a current
carrying wire : Linear colloidal superstructures are expected to follow
imposed magnetic field lines. Thus, by controlling field lines we control the
cluster geometry itself. A simple current carrying wire and the Biot-Savart
circular field structure around them appear as a natural candidate to form
rings. However along this avenue, once again, we quickly meet the roadblock
posed by kinetic traps in the complex space of colloidal cluster configurations.

In this work we report on a method to overcome this kinetic
problem by introducing what we call a "\textit{magnetic
merry-go-round field"} - a rotating field, dynamically modulating
the Biot-Savart field - around the wire axis that effectively
stretches the amorphous clusters into rings and helices. In the
first part of the paper we present the basic geometry of the setup
and the induced dynamic fields, as well as their effects on
cluster reshaping. In a second part we theoretically rationalize
for the experimentally observed cluster reshaping behavior by
developing a simplified, analytically tractable dimer toy-model.
We conclude with an outlook giving a glimpse of further
possibilities of the wire trap and ideas that could be explored in
the future.

\begin{figure*}[t]
\includegraphics[width=0.98\linewidth]{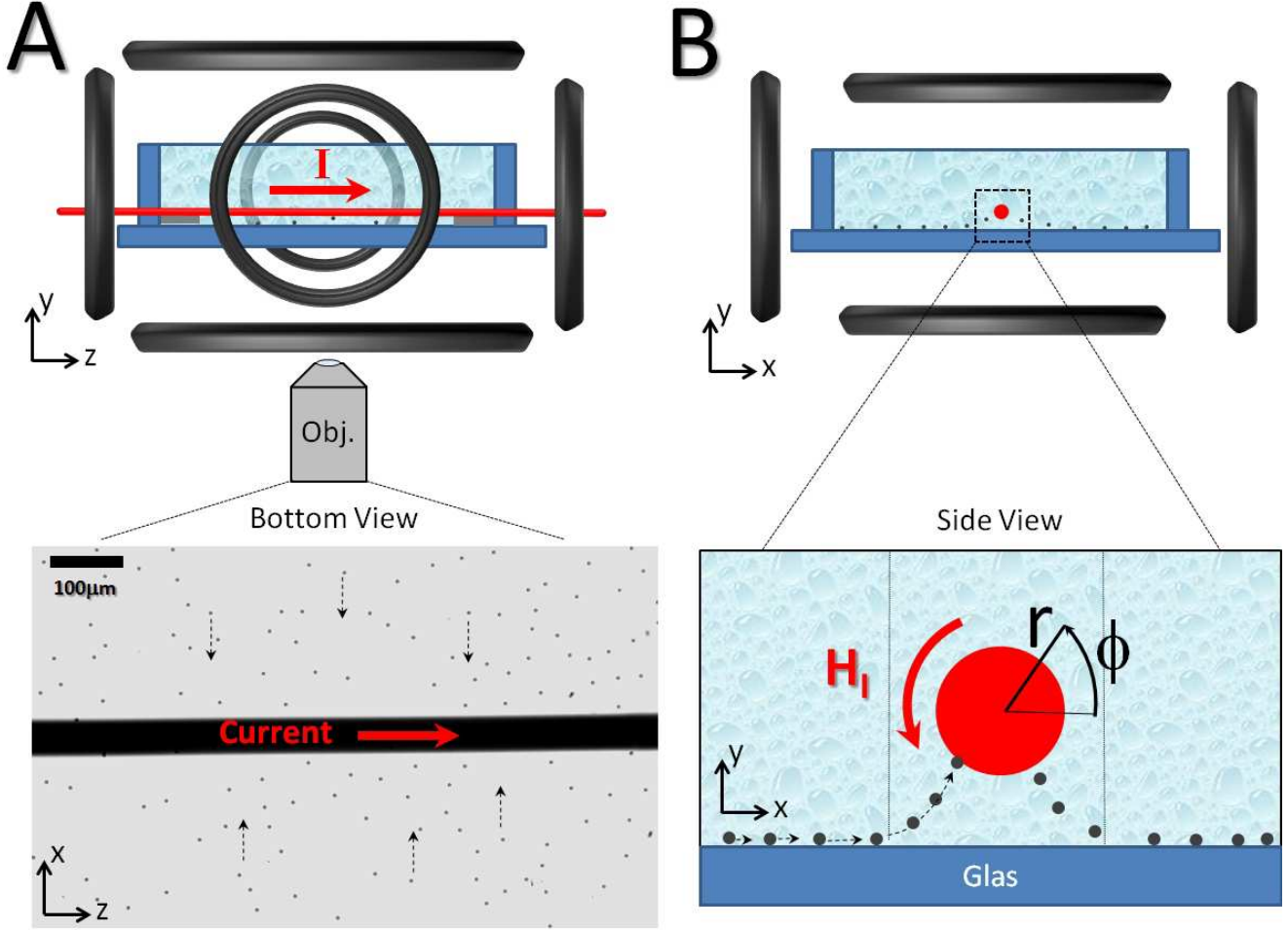}
\caption{ The wire trap experimental setup : A $50\mu m$ current carrying
copper wire suspended over a glass substrate traps superparamagnetic beads.
Additional pairs Helmholtz coils induce uniform fields in the x, y and z
direction that modulate the field of the wire. }%
\label{fig1}%
\end{figure*}

\section{Materials and Methods}

\textit{Superparamagnetic Beads.} The experiments were performed
with superparamagnetic beads consisting of a polystyrene polymer
matrix doped with magnetite nano-crystals (Dynabeads M-450,
Invitrogen). In order to prevent aggregation and surface adhesion,
the superparamagnetic particles were suspended in a $5.9
\times10^{-3}~\mathrm{M}$ sodium dodecyl sulfate solution
\cite{KreuterThesis}. The beads have a diameter $d_{b}(\equiv2
r_{b}) = 4.4 \mu m$, a mass density $\rho_{b} = 1.5 \times10^{3}
kg/m^{3}$ and the magnetic bead susceptibility $\tilde{\chi_{b}} =
6.2 \times10^{-11} Am^{2}/T$, \cite{Maret1}. The latter is related
to the dimensionless bead susceptibility $\chi_{b} = \mu_{0}
\tilde{\chi_{b}}/V_{b} = 1.75$, with $\mu_{0} = 4 \times10^{-7}
Tm/A$ the magnetic permeability of vacuum and $V_{b} =\frac{\pi
}{6}d_{b}^{3}$ the beads volume. Susceptibilities are defined as a
response function of the magnetization M or the magnetic moment $m
= M V_{b}$ to the external magnetic field $m = \tilde{\chi_{b}}
B_{e} = \chi_{b} V_{b} H_{e}$ and $M = \chi_{b} H_{e}$. In the
following we omit the index in $B_{e}$ and $H_{e}$.

\textit{Bead trapping on the wire - } The colloidal suspension is placed
inside a silicone chamber (Coverwell perfusion chamber, height 1~mm) that was
glued to a microscope glass slide. Due to their large density the colloidal
beads sediment on the microscope glass slide. An uninsulated copper wire
(diameter $d_{w} (\equiv2 r_{w}) =50~\mu m$ connected to a constant current
generator passes through the experimental cell, as sketched in Fig. (1).
Spacers of height $\approx100~\mu m$ ensured that the wire is suspended above
the microscope slide surface, and its height $h$ was measured before each
experiment. In order to generate magnetic fields able to attract particles at
the surface of the wire, significant currents must go through the wire,
generating high current densities up to $200-300~A/mm^{2}$, giving rise to
fields of several $mT$-at the wire surface). The thin wire could support
currents up to 0.5 A for several minutes and up to 0.8 A for short times
without deterioration.

The high currents through such thin wires can be achieved due to a more
efficient heat dissipation on the surface of micrometric wires in aqueous
solutions. The Joule-heating power per length of a wire of radius $r_{w}$
scales as $P_{J}/l\sim\rho r_{w}^{-2}I^{2}$ with $\rho$ the specific
resistivity, while the dissipated power due to advection cooling $P_{C}/l\sim
r_{w}h\Delta T$ \ with $h$ the heat transfer coefficient and $\Delta T$ the
temperature difference between the wire and the solvent (in the bulk). To
maintain a small limited temperature difference $\Delta T$ , the power balance
$P_{J}$ $=P_{C}$ implies a maximal current scaling as $I^{2}\sim\rho_{w}%
^{-1}r^{3}h\Delta T.$

Additional fields generated by a set of commercial and custom made Helmholtz
coils along the x, y or the z direction were superimposed with the field of
the wire. The coils were driven by a computer controlled signal generator
linked to a custom amplifier that allows to choose the amplitudes and phase
differences of the magnetic fields in the frequency range of 0.1 to 20 Hz.
Observation of the beads was performed on an inverted microscope, Eclipse Ti-S
(Nikon), equipped with a PlanFluor 10x/0.30 objective. Images were time lapse
recorded with a CCD camera (Hamamatsu) and subject to image analysis in
NIH-ImageJ - a Java-based, open source software package.

\section{Trapping beads on the wire, forming helices and rings.}

A simple method to generate circularly closed or helical field lines is a
current carrying wire. According to the Biot-Savart law, the wire forms closed
circular field lines around itself and should act as an elegant "field mould"
for colloidal particle rings around its circumference. In general , a wire
carrying a current $I$\ gives rise to a magnetic field at the position\textbf{
}$\mathbf{x}$ given by $\mathbf{H}_{I}\left(  \mathbf{x}\right)  =\frac
{I}{4\pi}\int\frac{d\mathbf{x}_{w}}{ds}\times\frac{\left(  \mathbf{x}%
-\mathbf{x}_{w}\right)  }{\left\Vert \mathbf{x}-\mathbf{x}_{w}\right\Vert
^{3}}ds$ , where $\mathbf{x}_{w}\left(  s\right)  $ is the centerline position
of the wire, $s$ its arc-length and $\mathbf{t}=d\mathbf{x}_{w}/ds$ the wire's
centerline tangent vector. In the simplest case of a straight long wire the
expression simplifies to $\mathbf{H}_{I}\left(  \mathbf{x}\right)  =(I/2\pi
r)\mathbf{e}_{\phi}$ with\ $r=\left\vert \mathbf{x}\right\vert $ the radial
distance to the wire center, $\phi$ the azimuthal angle and $\mathbf{e}_{\phi
}=\mathbf{e}_{y}\sin\phi-\mathbf{e}_{x}\cos\phi$ the azimuthal unit vector. In
general, a spherical, paramagnetic bead placed in an external magnetic field
$\mathbf{H}\left(  \mathbf{x}\right)  $ has the free energy given
by\textbf{\ }%
\begin{equation}
W(\mathbf{x})=-\frac{\mu_{0}}{2}\mathbf{m}(\mathbf{x})\mathbf{H}\left(
\mathbf{x}\right)  =\mathbf{-}\frac{\mu_{0}}{2}\mathbf{\chi}_{b}%
V_{b}\mathbf{H}^{2}\left(  \mathbf{x}\right)  ,\label{F0}%
\end{equation}
where $\mathbf{m}(\mathbf{x})\equiv\chi_{b}V_{b}\mathbf{H}(\mathbf{x})$ the
magnetic moment of the bead , $\chi_{b}$ is the bead's susceptibility and
$V_{b}$ its volume. The bead experiences a magnetic gradient force%
\begin{equation}
\mathbf{F}=-\partial W/\partial\mathbf{x}=\frac{\mu_{0}\chi_{b}V_{b}}{2}%
\nabla\left(  \mathbf{H}^{2}\left(  \mathbf{x}\right)  \right)  .\label{FI}%
\end{equation}
\textbf{\ }

We have built a simple setup consisting of a $50$ $\mu m$ (radius $25\mu
m$)\ thick long slender copper wire suspended above a glass surface, see
Fig.\ref{fig1}. After switching-on the current through the wire,
superparamagnetic beads that previously sedimented onto the bottom surface by
gravity interact with the field, Eq.\ref{FI}, and begin now to move towards
the wire. They are attracted to the wire with a force acting radially to the
wire (in $\mathbf{e}_{r}$ direction) given by%

\[
\mathbf{F}=-\frac{\mu_{0}\chi_{b}V_{b}}{4\pi^{2}}\frac{I^{2}}{r^{3}}%
\mathbf{e}_{r}%
\]

If the wire is sufficiently close to the substrate plane (typically
$<100-150\mu m$) and the current is high enough ($I>100$ $mA$) the beads that
have reached a close proximity to the wire begin to lift off the glass surface
and attach to the bottom of the wire. This behavior sets in once the magnetic
force overcomes the beads gravity force $F_{g}=\left(  \rho_{b}-\rho
_{sol}\right)  gV_{b}\approx0.22pN$ with $\left(  \rho_{b}-\rho_{sol}\right)
\approx0.5g/ml$ the bead-solvent density contrast and $g=9.8\frac{m}{s^{2}}$
the gravity acceleration. Once they lift to the wire, the beads can be held on
the wire with much smaller currents of the order $I=2\pi\left(  F_{g}r_{w}%
^{3}/\mu_{0}\chi_{b}V_{b}\right)  ^{1/2}\approx35mA$.

On the wire surface the beads form two dimensional clusters and short chains
that progressively grow over time as a result of their dipole-dipole
interactions. The dipole-dipole interactions tend to align the longer axes of
chain-like clusters along the field lines, which in the case of simplest wire
current field are pointing around the azimuthal direction of the cylindrical surface.

\begin{figure}[t]
\includegraphics[width=0.98\linewidth]{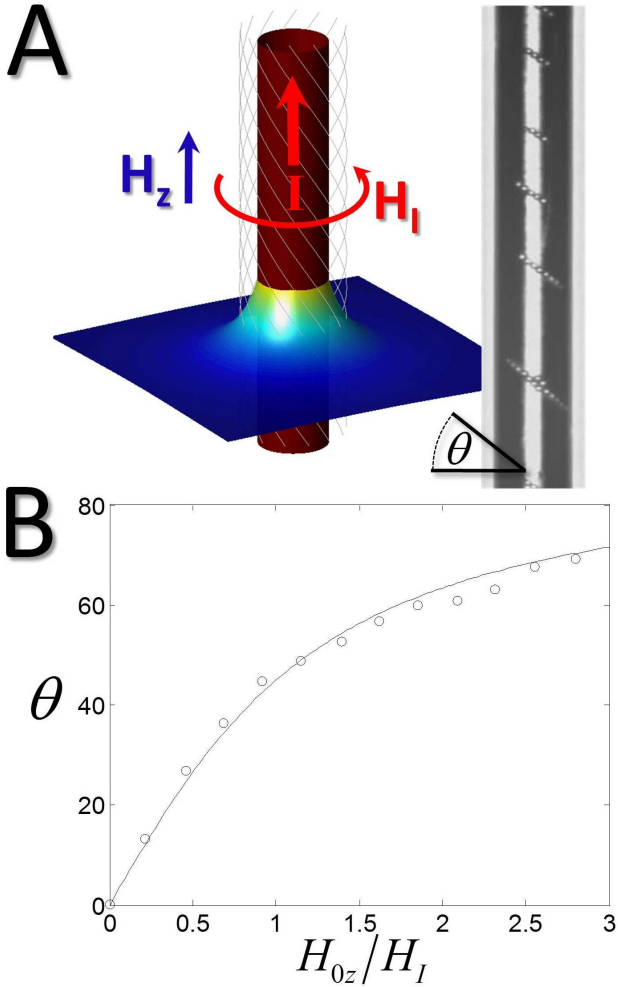}
\caption{ A) A combination of the wire field and a z directed
field induces helical field-lines (the total intensity $H^2$ is
color coded) that guide the formation of circular arcs and helical
filaments of varying pitch. B) The helix pitch as a function of
external field
over the wire field (circles) and the theoretical value (line).}%
\label{fig2}%
\end{figure}

The shape and alignment of field lines and in turn the alignment of the
clusters can be manipulated by invoking an additional external field
$\mathbf{H}_{0}$\textbf{\ }that superimposes with the wire field to give a
total external field $\mathbf{H}\left(  \mathbf{x}\right)  \mathbf{=\mathbf{H}%
_{0}+H}_{I}\left(  \mathbf{x}\right)  $. In the first and simplest case, the
uniform field points \textit{parallel} to the wire axis $\mathbf{H}_{0}%
=H_{0z}\mathbf{e}_{z}$. The total field in this case is given by%
\begin{equation}
\mathbf{H}\left(  r\right)  =H_{I}\left(  r\right)  \mathbf{e}_{\phi}%
+H_{0z}\mathbf{e}_{z} \label{helicalField}%
\end{equation}
with $H_{I}\left(  r\right)  =I/2\pi r$. Superimposed, these two fields\ form
helical field lines, with a pitch angle $\theta=\arctan\left(  H_{0z}%
/H_{I}\right)  $ that is\ followed by colloidal chains, see
Fig.\ref{fig2}.

By changing the proportion of the current $I\ $and the external field
intensity $H_{0z}$ one can manipulate the pitch angle $\theta$ of the helices,
see Fig. 2. Due to the orthogonality of the two component fields , $i.e.$
$\ \mathbf{e}_{\phi}\cdot\mathbf{e}_{z}=0$, the energy (up to a constant
term)\ and the gradient force resulting from Eq.(\ref{helicalField}) and
Eq.(\ref{FI}) remain unchanged w.r.t. the pure wire field case ($H_{0z}=0$).
That is, while chains become reoriented by the presence of the $z-field$,
single particles experience no change of trapping force or distribution on the surface.

\begin{figure}[t]
\includegraphics[width=0.98\linewidth]{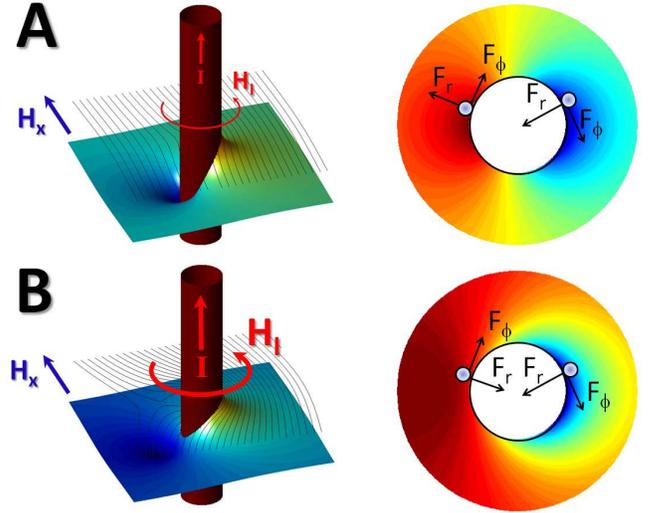}
\caption{ The "magnetic merry-go-round trap" field configuration: A
perpendicular field $H_{x}$ superimposed to the wire field $H_{I}$ generates a
complex trapping field geometry. Left column:\ the total squared field
strength $H^{2}$ , color code blue(red):\ low(high) H$^{2}$ . Right
column:\ Potential energy of a superparamagnetic bead (color code \ blue
:\ low energy) A)\ For a strong enough $H_{x}>H_{I}$ the normal component of
the magnetic gradient force changes sign and is radially repulsive on one
side. B)\ For sufficiently week external field $H_{x}<H_{I}\ $the trapping
field is everywhere attractive towards the wire surface. The beads remain on
the surface but experience azimuthal forces towards a single minimum energy
position. }%
\label{fig3}%
\end{figure}

In a second, more interesting case we impose an external field
\textit{orthogonal} to the wire axis (e.g. in the $x$ direction)%

\begin{equation}
\mathbf{H}\left(  r,\phi\right)  =H_{I}\left(  r\right)  \mathbf{e}_{\phi
}+H_{0x}\mathbf{e}_{x}. \label{orthoField}%
\end{equation}
Note, that now the field combination loses azimuthal symmetry and depends both
on the radial distance $r$ and the azimuthal angle $\phi$,\ see Fig.\ref{fig3}%
\textbf{.} A single bead in this field has the free energy%

\begin{equation}
W\left(  \phi,r\right)  =-\frac{\chi_{b}\mu_{0}V_{b}}{2}\left(  \frac{I^{2}%
}{4\pi^{2}r^{2}}-\frac{H_{0x}I}{\pi r}\cos\phi\right)  +const.
\label{W_SpinningField}%
\end{equation}
From the azimuthal component $F_{\phi}=-\left(  \partial W/\partial
\phi\right)  /r$ of the gradient force%

\begin{equation}
F_{\phi}=\frac{\chi_{b}\mu_{0}V_{b}}{2\pi}\frac{H_{0x}I}{r^{2}}\sin
\phi\label{asimForce}%
\end{equation}
and its radial component $F_{r}=-\left(  \partial W/\partial r\right)  $%
\begin{equation}
F_{r}=-\frac{\chi_{b}\mu_{0}V_{b}}{4\pi^{2}}\frac{I^{2}}{r^{3}}\left(
1-\frac{2\pi H_{0x}r}{I}\cdot\cos\phi\right)  \label{radForce}%
\end{equation}
we see that depending on the magnitude of $H_{0x}$ vs $I\ $and the angular
position $\phi$ the wire surface can be either\textit{ attractive} or
\textit{repulsive} in the radial direction. In the following we will focus
entirely on the weak external field, $H_{0x}<\frac{I}{2\pi r}$ , where the
wire is radially attractive\cite{NOTE1}$,$ i.e. $F_{r}<0,$ and the beads are
pressed onto the wire for any $\phi$ as in the case $H_{0x}=0$. The presence
of the symmetry breaking term $H_{0x}\mathbf{e}_{x}$ generates a preferred
orientation for the bead on the wire surface, i.e. a free energy minimum at
$\phi=\pi.$ This induced energy minimum and the fact that beads as well as
clusters follow it, will be used in the following to dynamically manipulate
and transform clusters.

\section{Dynamical Transformations by Helical Fields and Spinning Waves}

The wire trap appears to be an elegant tool to form helices and rings on the
wire, however the method meets practical some obstacles. Attracting beads to
the wire from the bottom surface leads mostly to the formation of random,
kinetically trapped, bulky clusters on the bottom side of the wire. Forming
defect-free chain and ring structures via this process seems therefore very
difficult and an unlikely process to happen spontaneously.

To remedy this problem our intuitive approach consists of applying additional
\textit{dynamic, time-dependent} fields to generate mixing of beads on the
wire surface and transform them into linear chain configurations.

\begin{figure}[t]
\includegraphics[width=0.8\linewidth]{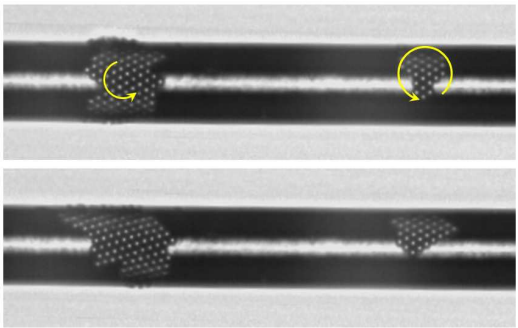}
\caption{ Helically spinning fields lead to the formation of hexagonal
clusters that spin with the field. }%
\label{fig4}%
\end{figure}

We have explored two different forms of dynamical external field
perturbations:\ $(a)$\ Rotating helical field and $(b)$\ azimuthal spinning waves.

$(a)$\textit{\ Rotary helical fields}, generated by an oscillating wire
current $I\propto\sin\omega t\ $ and an out of phase longitudinal field
$H_{0z}\propto\cos\omega t$ .

In this case, at each spot of the wire the field is tangential and rotates
around the cylindrical wire surface normal . We observe that cluster start to
rotate, collide and form hexagonally ordered bigger clusters. At higher bead
surface densities these rotating and growing clusters eventually cover the
surface of the wire with a colloidal mono-layer. While this behavior is
interesting in itself it does not lead to the desired elongation of clusters
into chains.

$(b)$ \textit{Azimuthal spinning waves ("Magnetic merry-go-round")} - They are
generated by a constant $I$ and a \textit{perpendicular oscillating external
field }of the form:%
\begin{equation}
\mathbf{H}_{0\perp}(t)=\mathbf{e}_{x}H_{0}\sin\omega t+\mathbf{e}_{y}H_{0}%
\cos\omega t. \label{perpField}%
\end{equation}
\textit{ }

\begin{figure*}[t]
\includegraphics[width=0.98\linewidth]{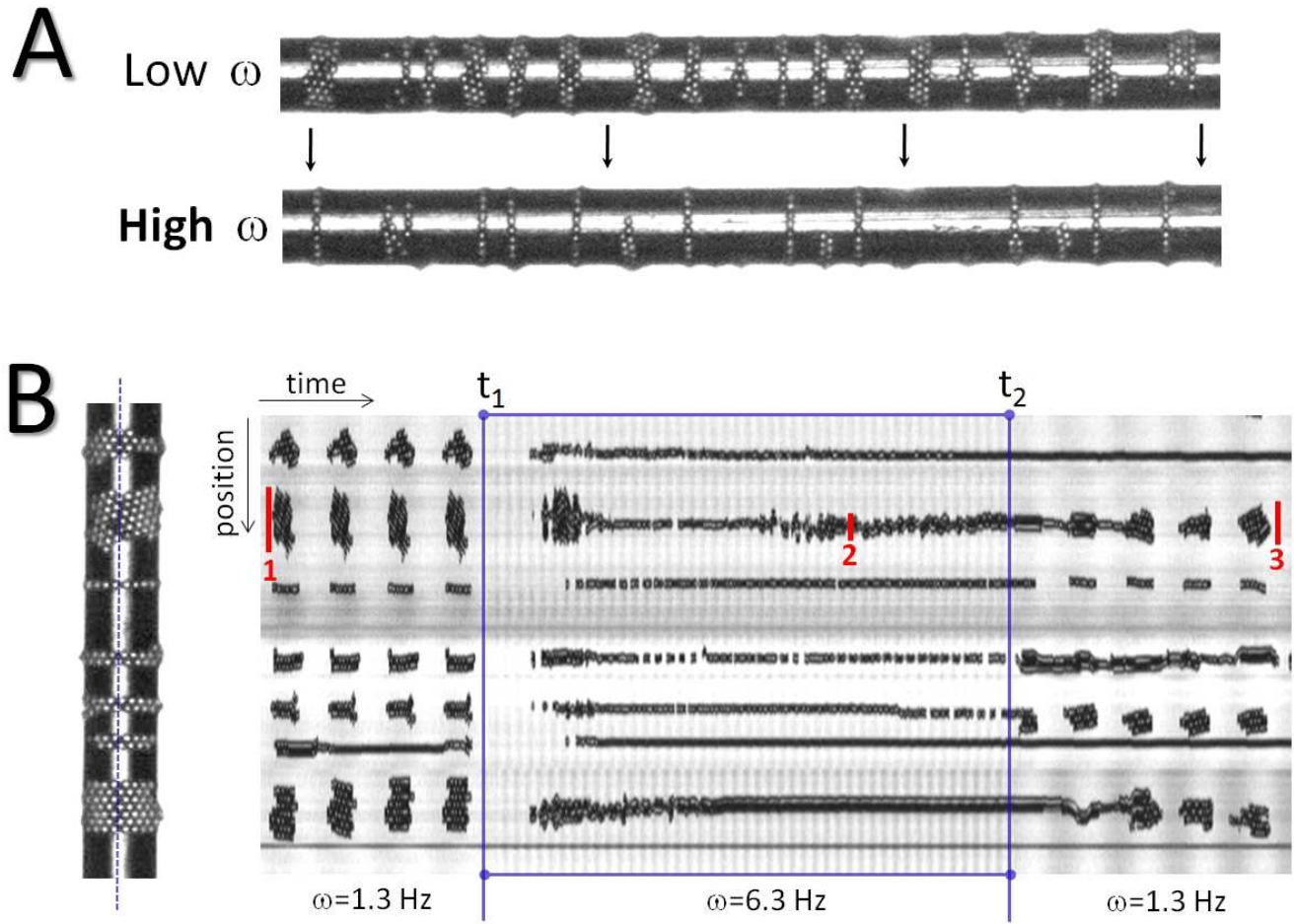}
\caption{ Transformations under the merry-go-round, azimuthally
spinning field: A) Amorphous clusters transform into elongated
chains and rings upon the activation of the spinning field. B) A
kymograph of the reshaping dynamics of several transforming
clusters under the spinning field at various frequencies. The time
interval of higher-frequency field spinning, between time $t_{1}$
and $t_{2}$ (blue box) the excitation leads to a lateral thinning
and longitudinal elongation of the clusters. Reverting to a lower
frequency spinning partially reverts the cluster elongation. The
lateral width of a
typical cluster at three different times is indicated by red lines.}%
\label{fig5}%
\end{figure*}

In this case of rotating fields orthogonal to the wire the dynamical behavior
is very much richer than in case $(a)$. The fields of the wire and external
field, as well as the induced dipolar fields between beads, dynamically
interfere in an interesting manner. The perpendicular field $\mathbf{H}%
_{0\perp}(t)$ superimposes with the wire field $\mathbf{H}_{I}\left(
\mathbf{x}\right)  =\frac{I}{2\pi\left\vert \mathbf{x}\right\vert }%
\mathbf{e}_{\phi}$ and weakens it on one side but strengthens it on the other
side of the wire leading to the formation of a wave-like field structure on
the surface of the wire with a free energy of the form $W\left(  \phi-\omega
t,r\right)  $ (Eq. \ref{W_SpinningField} )- ,\ see Fig.\ref{fig3} and the
previous section for the static case\textbf{.}

\begin{figure}[t]
\includegraphics[width=1.0\linewidth]{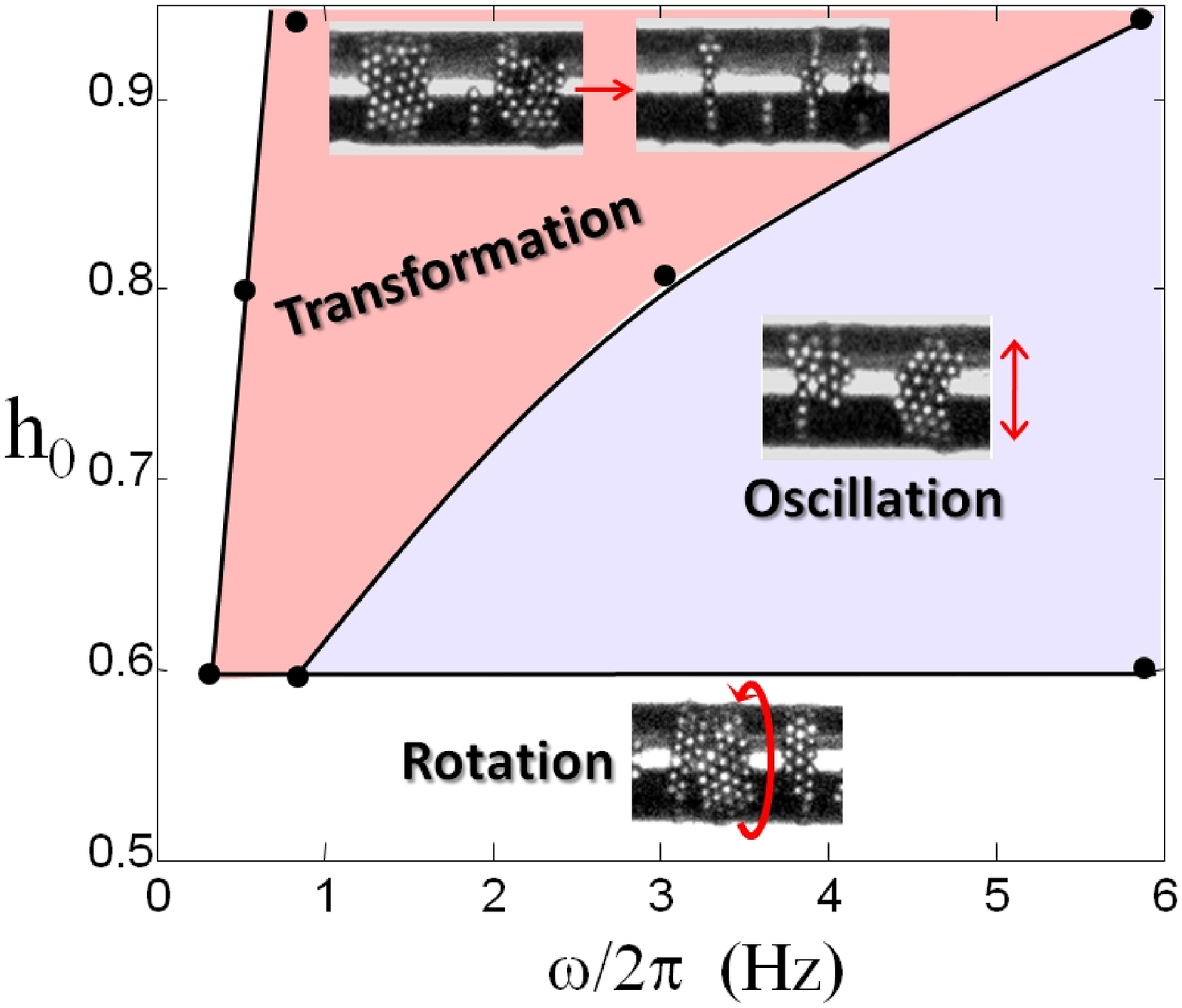}
\caption{ Phase diagram of clusters in a spinning wave field as a function of
the scaled external field $h_{0}=\frac{H_{0}}{H_{I}}$ and the field spinning
angular frequency $\omega$. }%
\label{fig6}%
\end{figure}

In the following we shall investigate how\textbf{\ }the
propagation of spinning "merry-go-round" waves around the wire
axis leads to a complex dynamical phenomenology, see Figs
\ref{fig5},\ref{fig6}. The behavior of clusters in spinning fields
depends on two parameters : the field \textit{spinning angular
frequency} $\omega$ and the \textit{relative field intensity}
$h_{0}=\frac{H_{0}}{H_{I}}$.\textbf{\ }We find that when the
frequency of the spinning wave is sufficiently small
$\omega<\omega_{c}$, with $\omega_{c}$ a critical frequency, one
observes clusters of beads rotating uniformly with the field
around the wire without a change of shape. In this "field locked"\
regime of slow field spinning, the clusters are following the
field-maximum (i.e. the free energy minimum) on the wire surface
with the same frequency but with a small constant angular lag. The
clusters' angular lag behind the free energy minimum grows with
increasing frequencies . From a critical frequency
$\omega=\omega_{c}$ the clusters start to fall behind the
potential more than a full turn. In this "unlocked" or "field
skipping"\ regime, the clusters move with their own cluster
frequency which is lower than the field frequency. During this
motion the clusters occasionally skip the free energy maximum
along their trajectory that leads them from one free energy
minimum to the next one. For such supercritical frequencies
$\omega\gtrsim\omega_{c}$ and at the same time large enough
orthogonal field strengths, $h_{0}\approx1,$ the clusters start to
experience strong stretching forces at the free energy maxima.
They in turn elongate and transform into slender, ring like
structures. This transformation behavior becomes more frequent
with growing frequency. However, with further increasing frequency
the transformations gradually start to diminish for very rapid
fields. Finally, for very high frequencies $\omega\gg\omega_{c}$
the spinning field becomes too fast and the clusters stop
following the field dynamics. In the very fast field spinning
regime the field does not affect the cluster shape significantly.
Apart from small oscillations and a very slow azimuthal net
rotation the clusters do not transform in this fast driving limit.

An example of the reshaping dynamics of transforming clusters at
various frequencies is shown in Fig.\ref{fig5} and the diagram of
transformation behavior in Fig.\ref{fig6}. Interestingly switching
from high frequencies back to lower ones, for sufficiently long
time periods, can partially revert the effect of chain elongation,
giving rise to crumpling and squeezing of clusters back into a
more compact shape. This, in our case undesired, reverse buckling
process can effectively be suppressed by abruptly stopping the
spinning field (while maintaining the current in the wire
constant), thus leaving the chain statically "frozen" in the
desired elongated state.

\section{Origin of Cluster to Chain Shape Transformation}

The phenomenon of dynamic cluster elongation into chains is rather
remarkable and deserves a theoretical investigation. We present it
in the following Section for a simplified "dimer-toy model".
Central to the model is the interplay of two dynamic phenomena
that occur during the propagation of spinning waves around the
wire. The first phenomenon for cluster transformation is the
\textit{locking /\ unlocking transition} of the cluster with
respect to the rotating field. An elongating cluster should become
unlocked and skip over the field barrier in order to become
transformed. This transient event can lead to stretching and
reshaping of the cluster. The second requirement for
transformation is a \textit{sufficiently high external field} at
the point of barrier crossing. That is, under certain conditions,
that will be clarified below, the crossing of the energy barrier
involves a sufficiently destabilizing field able of tearing apart
and rejoining the cluster into an elongated form.

During the cycle of potential skipping, the cluster experiences
both, tension (at the potential maximum) and compression (at the
potential minimum). This leads to the naive idea that elongation
should be reverted, i.e. when the cluster elongates during the
tension period it should reshape back analogously into the more
compact form during the compression phase. This reversibility is
however prohibited by a third phenomenon that we call
\textit{configurational hysteresis}. As we show in the Appendix
II, a magnetic chain with a side defect that is exposed to a
cycling tension and compression will undergo a hysteresis in its
shape. During the tension period the defect of the chain becomes
unstable, smooths out and elongates but in the compression period
it behaves asymmetrically. In particular, under appropriate
dynamic conditions, the chain does not buckle back into the
initial shape due to different barriers in both directions.
Overall this asymmetric behavior tends to favor the elongation of
magnetic clusters in general.

\subsection{Dimer Toy-Model for Cluster Transformation}

Here the simplest analytically tractable model for cluster rearrangement - a
simple two bead \textit{dimer toy-model}\textbf{.} Of course, this simplest
"proto" cluster cannot rearrange irreversibly beyond the two beads being
simply pulled apart. However, the event of parts of the cluster coming apart
is a central ingredient in all cluster transformations. Following the
stretch-out and mutual separation of two sub-clusters, side chain beads pop-in
and fill the gap between. In this sense studying the dimer toy-model has all
of the ingredients for the precursor event of transformation - namely the
dimer opening. With this interpretation of opening as a precursor, the dimer
system mimics the phase-diagram of larger cluster transformation qualitatively
rather well.

\begin{figure}[ptb]
\includegraphics[width=0.6\linewidth]{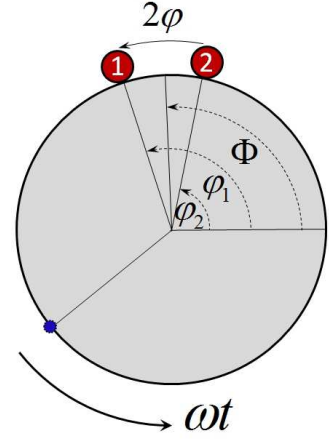}
\caption{ The bead dimer model: The angular center of mass $\Phi$ of beads 1
and 2 follows the free energy minimum (blue dot) that is rotating with an
angular velocity $\omega.$\ The dimer's relative coordinate $\varphi$ is an
indicator for its opening and reshaping dynamics. }%
\label{fig7a}%
\end{figure}

Consider two beads sitting on the wire (with the radius $r_{W}$) and acted
upon the azimuthally spinning external field\textit{\ }$\mathbf{H}%
_{0}(t)=\mathbf{H}_{0\perp}(\mathbf{x},t)=\mathbf{e}_{x}H_{0}\sin\omega
t+\mathbf{e}_{y}H_{0}\cos\omega t$, with bead coordinates $\mathbf{x}%
_{1,2}=(r_{w},\varphi_{1,2})$ - see the geometry in Fig. \ref{fig7a}. \ The
two colloids interact with the external field and with each other via induced
dipolar forces.The corresponding total free energy can be written as $W\left(
\varphi_{1,2},t\right)  =W_{0}+W_{dip}+W_{ev}$\ with $W_{0}$ the interaction
energy with the field given in $Eq.$ (\ref{F0}), $W_{dip}$ the the
dipole-dipole interaction and $W_{ev}\left(  \varphi_{1}-\varphi_{2}\right)  $
an excluded volume, short range, repulsive interaction between the
beads\textbf{.} \ The total free energy $W$ can be conveniently expressed as a
function of\textit{\ the center of mass} $\Phi=(\varphi_{1}+\varphi_{2})/2$
and the \textit{relative coordinate} $\varphi=(\varphi_{1}-\varphi_{2})/2$ .

In the limit of slow, viscosity-dominated motion (i.e. at zero Reynolds
number) the dynamics is governed by the Rayleigh dissipation function
$R_{diss}=\frac{1}{2}\xi\left(  \dot{\varphi}_{1}^{2}+\dot{\varphi}_{1}%
^{2}\right)  $. The friction coefficient $\xi\propto
r_{w}^{2}\eta$ is a phenomenological parameter that includes not
only the solvent (dynamic)\ viscosity $\eta$, but also lubrication
effects between the beads and the wire surface. For the
calculations it is useful to introduce the center of mass angle
coordinate in the field co-moving system $\Theta =\Phi+\omega t.$
The dynamic equations of the two beads are given by $\partial
R_{diss}/\partial\dot{\varphi}=-\partial W/\partial\varphi$ and
$\partial R_{diss}/\partial\dot{\Theta}=-\partial
W/\partial\Theta$ and explicitly written in $Eqs.$
(\ref{phi-dot}-\ref{Theta-dot}) of $Appendix$ $I$. In the limiting
case of $r_{b}/r_{w}\ll1$ and $\varphi\gtrapprox
\varphi_{\min}\approx r_{b}/r_{w} $ can be simplified to:%
\begin{equation}
\dot{\Theta}=\omega+\omega_{c}\sin\Theta\label{JJ}%
\end{equation}

\begin{equation}
\alpha\dot{\varphi}=-\frac{3}{2}\left[  h_{0}^{2}+4\right]  +2h_{0}\cos
\Theta-\frac{9}{2}h_{0}^{2}\cos2\Theta. \label{apr-phi-dot}%
\end{equation}

Here $\omega_{c}=2h_{0}/\alpha$ is a characteristic frequency ,
and $\alpha=2\xi/\mu_{0}\chi_{b}H_{I}^{2}V_{b}$ a characteristic
timescale. By defining the effective "tilted wash-board" potential
$U(\Theta)=-\omega \Theta+\omega_{c}\cos\Theta$ the co-moving
center of mass Eq. (\ref{JJ}) can also be written as
$\dot{\Theta}=-\partial U/\partial\Theta$ \cite{NOTE2}.

\begin{figure*}[t]
\includegraphics[width=0.98\linewidth]{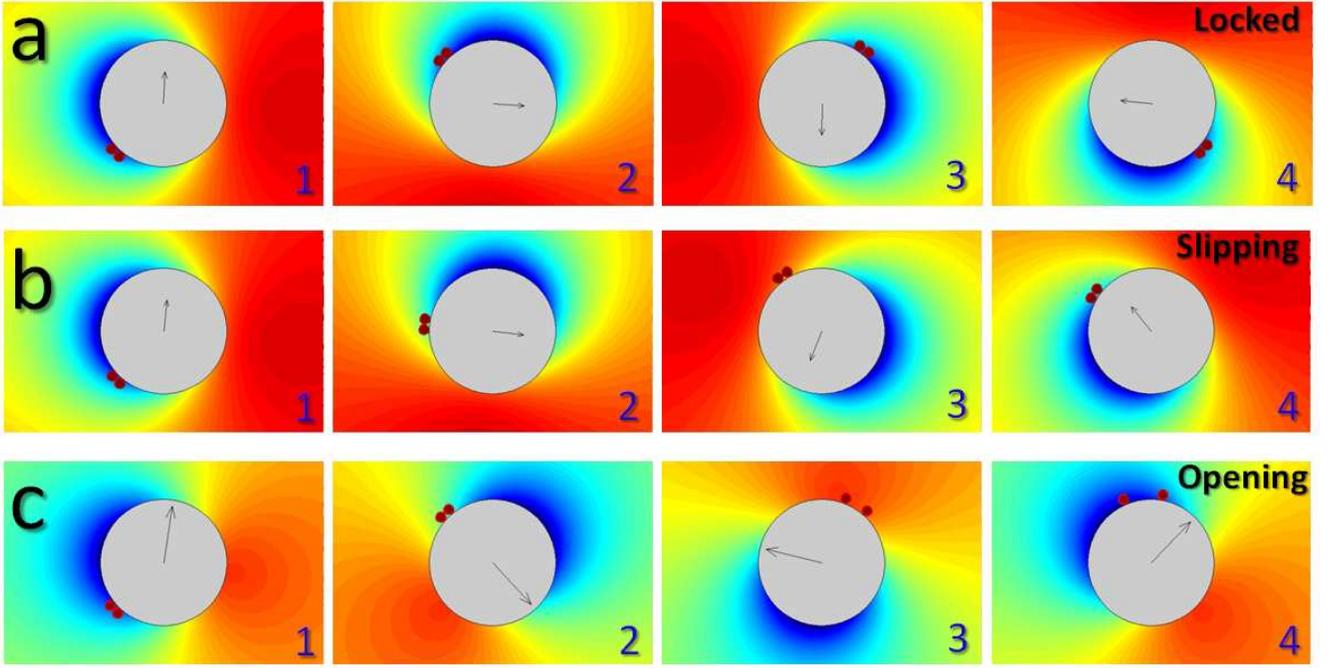}
\caption{ Snapshots from the simulation of a dimer under spinning
waves of various intensities and frequencies. The squared
intensity of the total field (wire field + external field) is
represented by the color code: high-field (low free energy) in
blue, low-field (high free energy) in red. The beads tend to
follow the free energy minimum (blue) for lower frequencies while
lagging behind it at higher frequencies. Grey arrow: the direction
of the external field. a)$h_{0}=0.5$ , $\omega=1$ : the dimer is
locked and moves with the field. b) $h_{0}=0.5$ , $\omega=2$ : the
field spins fast, the dimer lags behind the field but stays
closed. c) $h_{0}=0.9$, $\omega=3$: the dimer lags behind the
field and opens up while skipping the free
energy maximum.}%
\label{fig7}%
\end{figure*}

\begin{figure}[t]
\includegraphics[width=0.9\linewidth]{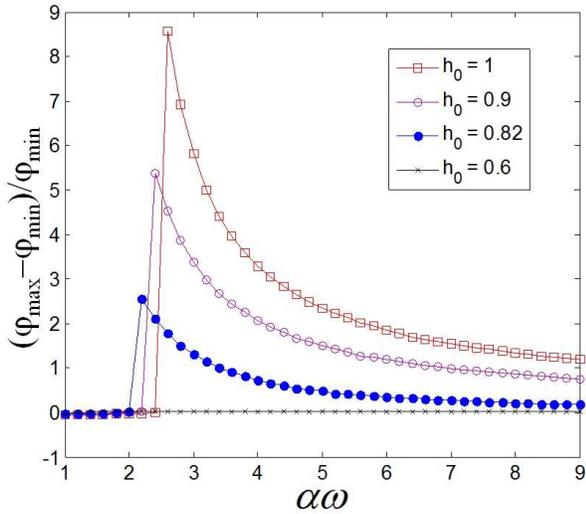}
\caption{The relative opening of the bead dimer for various $h_{0}$ as a
function of the dimensionless driving frequency.}%
\label{fig8}%
\end{figure}

\begin{figure}[t]
\includegraphics[width=0.98\linewidth]{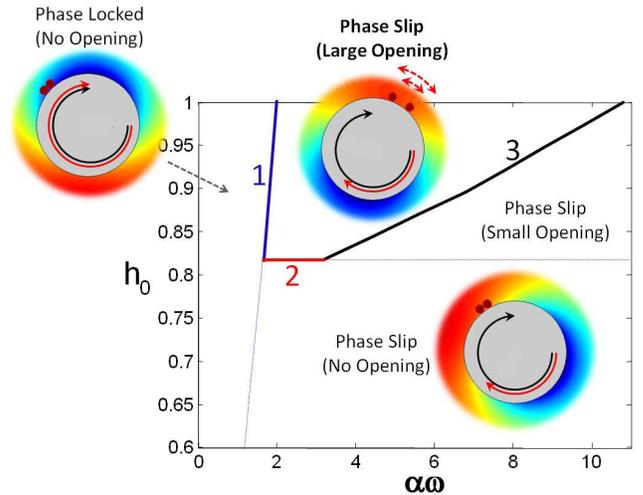}
\caption{ Phase diagram of a bead dimer on the wire obtained by numeric
integration of the equations of motion with an excluded volume term ensuring
bead non-interpenetration. Black/red circular lines stand for the angular velocity
of the field and the cluster respectively.}%
\label{fig9}%
\end{figure}

\bigskip

\subsection{Dimer Phase Diagram}

We obtained the numerical solution of the full set of equations
$Eqs.$ (\ref{phi-dot}-\ref{Theta-dot}) with an additional excluded
volume, hard core interaction at short range ,
$\varphi\leq\varphi_{\min}$. This allows us to study the opening
dynamics (see Figs. \ref{fig7}) and obtain the full phase diagram
(Fig.\ref{fig9}) of the dimer-toy model. The numerical
dimer-cluster solution, Fig \ref{fig7},\ref{fig8} shows two
regions of distinct dimer behavior that very much resemble the
experimental cluster phase diagram, Fig.\ref{fig6}. Within the
region delimited by three characteristic lines in Fig.
\ref{fig9}(1-3) the dimer opens sufficiently, i.e. order unity in
terms of the bead's diameter, while outside that region the dimer
does not open significantly. The position and shape of the
boundary lines can be understood analytically from a closer
analysis of Eqs.(\ref{JJ}) and (\ref{apr-phi-dot}) .

The \textit{line 1} (blue line , Fig.\ref{fig9})\ delimits the
region on the left where the dimer is field-locked and the region
on the right where the cluster lags behind the field. Its
approximate shape is obtained by studying the behavior of
solutions to co-moving center of mass coordinate, Eq. (\ref{JJ}).
These solutions are $\Theta\left(  t\right)  =cons.$ for
$\omega<\omega_{c}=2h_{0}/\alpha$ (i.e. the cluster is locked to
the field) and are unbounded for $\omega>\omega_{c}$ i.e. the
cluster starts to move relative to the field coordinate system.
Therefore the boundary line 1 is simply given by $h_{0}\left(
\omega\right)  =\alpha\omega/2.$

The second line, \textit{line 2 }(red line , Fig.\ref{fig9}), can
be understood by studying the relative coordinate Eq.
(\ref{apr-phi-dot}). Along this delimiting line, the sign of the
force acting on the relative coordinate flips from negative force
(tending to close the dimer) to positive force (with a tendency to
open it). The critical condition for line 2 is then given by
setting $\dot{\varphi}=0$ in Eq.(\ref{apr-phi-dot}) leading to the
condition on
the center of mass coordinate%

\begin{equation}
\cos\Theta_{1,2}=\frac{4}{3h_{0}}\frac{1\pm\sqrt{\frac{3}{4}h_{0}^{2}-\frac
{1}{2}}}{2}. \label{cosTheta}%
\end{equation}

The requirement of a real solution implies $h_{0}\geq h_{0,c}=(2/3)^{1/2}%
\approx\allowbreak0.82.$ Thus for $h_{0}$ beyond $h_{0,c}$ the dimer can open
at some time point its cycle and $h_{0}=(2/3)^{1/2}$ defines the delimiting
line 2.

The \textit{line 3 }(black line , Fig.\ref{fig9}) is given by the
condition that the relative coordinate opens up sufficiently at
some time point during the cycle for some $t.$ The amount of
opening in the phenomenological model , is to some extent
arbitrary. In the phase diagram shown we chose a critical opening
condition for an opening of order unity , i.e. $\varphi\left(
t\right)  >\left(  3/2\right)  \varphi_{\min}.$ In this corner of
the phase diagram, the dimer relative coordinate makes excursions
from the closed state $\varphi\left(  t\right)  =\varphi_{\min}$
to some maximal value $\varphi\left(  t\right)  =\varphi_{\max}$.
The size of this maximal excursion value $\varphi_{\max}$ we can
estimate from the typical time interval $\Delta
t_{open}=t_{2}-t_{1}$ for which a positive, dimer opening force
$f$ acts on the relative coordinate. The opening time interval is
related to the maximal opening angle
$\Delta\varphi=(1/\alpha)\int_{t_{1}}^{t_{2}}f(\Theta
(t))dt\approx(1/2\alpha)\Delta t_{ope}$. Setting the opening angle
to twice the minimal angle $\Delta\varphi=2\varphi_{\min}$ (so
that one additional bead would fit in between the maximally
separated dimers), in the limit $\omega
\gg\omega_{c}$\ one estimates $\Delta t_{open}\approx(2\sqrt{2}h_{0}%
-1.6)/\omega.$ The boundary line resulting from that is $h_{0}\left(
\omega\right)  \approx(1.6+\alpha\omega\Delta\varphi)/2\sqrt{2}$\ - in a
satisfactory agrement with the line $3$ (black) in Fig.\ref{fig9}.

The region of the dimer toy-model phase diagram that is confined
between these three critical lines (1-3)\ is also the region of
interest for real large cluster transformation. To the left,
bottom or right of this region, the cluster either does not open
at all or does not open sufficiently for a transformation to
occur. Quantitatively, the dimer-toy model overestimates the
position of the line 2. In reality the clusters transform at a
lower external field $h_{0}^{\exp}\sim0.6$ as compared to the
dimer simulation result $h_{0}^{\dim}\sim0.82.$ Nevertheless, the
shape of the transformation region is qualitatively captured and
is rather similar to the experimental cluster transformation
diagram, Fig.\ref{fig6}, taken the simplicity and various
approximations of the dimer-toy model.

In summary, we have seen that the dimer-toy model can phenomenologically
explain the observed experimental observations. For a low field rotating
frequency $\omega<\omega_{c}$ the center of mass of a cluster rotates
uniformly and the beads are bound to stable clusters. For $\omega>\omega_{c}$
and a suitably large external field $h_{0}>h_{0,c}$ the clusters are lagging
behind the field and begin to tear apart. This tearing apart in the toy-model
translates into a cluster-to-ring transformation as any two neighboring beads
(or sub-clusters) in a bigger cluster would become stretched apart by the same
mechanism. This in turn gives rise to gaps and holes between the sub-clusters,
thus allowing lateral beads to pop-into the hole. If the time of stretching is
sufficiently large a bead from the lateral side have enough time to enter
between two stretched beads (or sub-clusters) with the tendency of forming
elongated structures like rings.

\section{\textit{Conclusion and Outlook}}

We have described a simple, yet robust method to trap and manipulate magnetic
beads on a wire. While from equilibrium considerations we might expect rings
and helical chains to form under such geometries in practice in a static field
they are kinetically, notoriously improbable to form.\ However, by dynamically
modulating the field of the wire with additional uniform, oscillatory external
fields we have generated field geometries and dissipative forces that lead to
a transformation of clusters into the desired linear structures. The method
consists of "shaking down" colloidal structures in their free energy landscape
via the trick of configurational hysteresis.

The generation of spinning magnetic waves has some similarities
with propagating magnetic fields in flat geometries along planar
magnetic garnets in rotating fields
\cite{TiernoRatchetingOnGarnet}. It allows to simply and
inexpensively generate fluxes of magnetic particles along
surfaces\textbf{,} a phenomenon that could have some interesting
ramifications in particular for particle transport. Exploiting
geometry of curved conductors of more complex shapes in
combination with dynamic external field\ modulation opens an
interesting playground for colloid world. Using for instance a
helical wire combined with a perpendicularly rotating field would
provide the spinning waves with an longitudinal component and
enable pumping of particles along the wire surface
\cite{InPreparation}.

The discussed system represents also a minimalistic way to
self-assemble and operate a \textit{magnetic micromotor}: the wire
being the stator and the colloidal cluster (or a single bead)
acting as a rotor. A further down scaling of the system appears
possible as thinner wires can more efficiently dissipate heat and
carry larger currents allowing for similar trapping down to micron
scales.

Finally, cross-linking the templated structures into self standing objects
appears a very promising route in colloidal assembly. Helically cross-linked
chains should become interesting magneto-responsive actuators when acted upon
by larger fields.

\textit{Acknowledgements.} We thank Christian Kreuter for kindly
providing us with beads and the members of the M3 team for
fruitful discussions. MLK thanks the members of the project
ON171005 of the Institute of Physics, Belgrade for useful
discussions.

\section{Appendix}

\subsection{Appendix I: The free-energy and dynamical equations for two beads}

The free-energy of two beads in an external field $\mathbf{H}\left(
\mathbf{x}\right)  $ is $W=W_{0}+W_{dip}$ where $W_{0}$ - given by
Eq.(\ref{F0}), is the energy of the non-interacting beads. The dipole-dipole
energy $W_{dip}$ in leading order with respect to $\mathbf{H}(\mathbf{x})$ is
given by%

\begin{equation}
W_{dip}\approx-\mu_{0}\chi_{b}V_{b}\mathbf{H}\left(  \mathbf{x}_{1}\right)
\hat{D}(\mathbf{x}_{1},\mathbf{x}_{2})\mathbf{H}\left(  \mathbf{x}_{2}\right)
.\label{dip-en}%
\end{equation}
Here, $\mathbf{H}\left(  \mathbf{x}\right)  =\mathbf{H}_{I}\left(
\mathbf{x}\right)  +\mathbf{H}_{0}(t)$, $\hat{D}(\mathbf{x}_{1},\mathbf{x}%
_{2})=\Psi_{12}(1-3\hat{N}_{12})$, $\Psi_{12}=\chi_{b}V_{b}/4\pi\left\vert
\mathbf{x}_{1}-\mathbf{x}_{2}\right\vert ^{3}$ and $\hat{N}_{12}%
=\mathbf{n}_{12}\otimes\mathbf{n}_{12}$, $\mathbf{n}_{12}=(\mathbf{x}%
_{1}-\mathbf{x}_{2})/\left\vert \mathbf{x}_{1}-\mathbf{x}_{2}\right\vert $.
Two beads on the wire are characterized by angles $\varphi_{1},\varphi_{2}$.
For further analysis it is convenient to use the relative coordinate
$\varphi=(\varphi_{1}-\varphi_{2})/2$ and the center of mass coordinate
$\Phi=(\varphi_{1}+\varphi_{2})/2$. The explicit expression for the
free-energy reads ($\Tilde{W}=W/(1/2)\mu_{0}V_{b}H_{I}^{2}$)
\begin{align*}
\Tilde{W}(\varphi,\Phi) &  =const+2h_{0}[1+2D(\varphi)]\cos\varphi\cos
(\Phi+\omega t)]\\
&  -\frac{3}{2}h_{0}^{2}D(\varphi)\cos2(\Phi+\omega t)
\end{align*}%
\begin{equation}
-\frac{3}{2}D(\varphi)[h_{0}^{2}+(3+\cos2\varphi)],\label{W}%
\end{equation}
where $D(\varphi)=D_{0}(\sin\varphi_{\min}/\sin\varphi)^{3}$, $D_{0}=\chi
_{b}V_{b}/4\pi d_{b}^{3}$ and the minimum contact angle of two beads is given
by $\sin\varphi_{\min}=d_{b}/d_{w}$. By introducing the angle $\Theta
=\Phi+\omega t$, that measures the mean coordinate in the field co-moving
frame, the equations for $\varphi$ and $\Theta$ read
\begin{align}
\alpha\dot{\varphi} &  =-D(\varphi)\left\{  \sin2\varphi+\frac{3}{2}\frac
{\cos\varphi}{\sin\varphi}\left[  h_{0}^{2}+3+\cos2\varphi\right]  \right\}
\label{phi-dot}\\
&  +2h_{0}\left\{  \left[  1+2D(\varphi)\right]  \sin\varphi+6D(\varphi
)\frac{\cos^{2}\varphi}{\sin\varphi}\right\}  \cos\Theta\nonumber\\
&  -\frac{9}{2}D(\varphi)\frac{\cos\varphi}{\sin\varphi}\cos2\Theta.\nonumber
\end{align}%
\begin{align}
\gamma(\dot{\Theta}-\omega) &  =2h_{0}\left[  1+2D(\varphi)\right]
\cos\varphi\sin\Theta\label{Theta-dot}\\
&  -3h_{0}^{2}D(\varphi)\sin2\Theta,\nonumber
\end{align}
where $\alpha=2\xi/\mu_{0}\chi_{b}H_{I}^{2}V_{b}$. Note, that the terms due to
dipole-dipole interaction are proportional to $D(\varphi)$. In order to study
the dynamics of the bead one should take into account its impenetrability,
what formally implements a condition on $\varphi$ that $\varphi>\varphi_{\min
}$, i.e. the relative angle $\varphi$\ must\textbf{\ }be always larger than
the contact angle of two beads $\sin\varphi_{\min}=r_{b}/r_{w}$. (In our
experiments one has $\varphi_{\min}\approx r_{b}/r_{w}\approx0.1$.) In the
numerical calculations this effect is described by introducing a strong
repulsive potential.

Since the dipole-dipole energy is proportional to
$D(\varphi)<D_{0}$, where for our dynabeads one has
$D_{0}\approx0.08$, then for $h_{0}\gg D_{0}$ one can neglect the
corresponding terms in $Eq.$ (\ref{Theta-dot}). Since in our
experiment $\varphi\gtrsim\varphi_{\min}\approx0.1$ -
$\varphi_{\min}$ is a minimal opening angle for the stretching
(and entering of the third bead between the two beads), then in
the first approximation one can retain in $Eq.$ (\ref{Theta-dot})
the term with $\cos\varphi(\approx1).$ The center of mass equation
$Eq.$ (\ref{Theta-dot}) is now reduced to the Eq.(\ref{JJ}) in the
main text. The equation for $\varphi(t)$ is also analogously
simplified for $h_{0}\gg D_{0}$ to Eq. (\ref{apr-phi-dot}).

The structure of solutions\textbf{\ }of $Eq.$ (\ref{JJ}) (which
holds for $\varphi_{\min}\approx D_{0}\approx0.1$) depends on the
dimensionless parameter $\kappa=\omega/\omega_{c}$ (with
$\omega_{c}=2h_{0}/\alpha$). For $\kappa<1$ the solution is
relaxation-like
\begin{align}
&  \ln\frac{1-\sqrt{1-\kappa^{2}}+\kappa\tan\frac{\Theta}{2}}{1+\sqrt
{1-\kappa^{2}}+\kappa\tan\frac{\Theta}{2}}\label{Om<Omc}\\
&  =\tau\sqrt{1-\kappa^{2}}+const,\nonumber
\end{align}
where $\tau=\omega_{c}t$. Whatever initial conditions are, for
large $\tau$ this solution goes to one of the minima
$\Theta_{\min}^{n}=(2n+1)\pi +\arcsin\kappa$ of $U(\Theta)$. This
solution is characterized by the time averaged\ value
$\left\langle \dot{\Theta}\right\rangle _{t}=0$. For $\kappa>1$
the solution is given by
\begin{align}
&  \arctan\left[  \frac{1}{\sqrt{\kappa^{2}-1}}\left(  1+\kappa\sqrt
{\frac{1-\cos\Theta}{1+\cos\Theta}}\right)  \right]  \label{Om>Omc}\\
&  =\frac{\sqrt{\kappa^{2}-1}}{2}\omega_{c}t+const.,\nonumber
\end{align}
and it is characterized by $\left\langle \dot{\Theta}\right\rangle _{t}%
=\sqrt{\omega^{2}-\omega_{c}^{2}}$, i.e. the "voltage-current"
dependence is non--linear \cite{Bulaevskii}.

In the main text it is shown that \textit{the opening regime is possible} if
the condition $h_{0c}<h_{0}<1$ is fulfilled. As it was discussed above, the
condition $h_{0}<1$ means that the beads are attracted to the wire if
$H_{0}<H_{I}$, while for $H_{0}>H_{I}$ some beads in the rings (or clusters)
are repelled from the wire and the self-assembly of rings is stopped. The
condition for $h_{0c}$ is obtained from $Eq.$ (\ref{phi-dot}) which describes
the dynamics of the relative angle $\varphi>$($\varphi_{\min}$) of two beads.

\subsection{Appendix II: "Push-pull" hysteresis: macroscopic magnet
chain example}

\begin{figure}[t]
\includegraphics[width=0.98\linewidth]{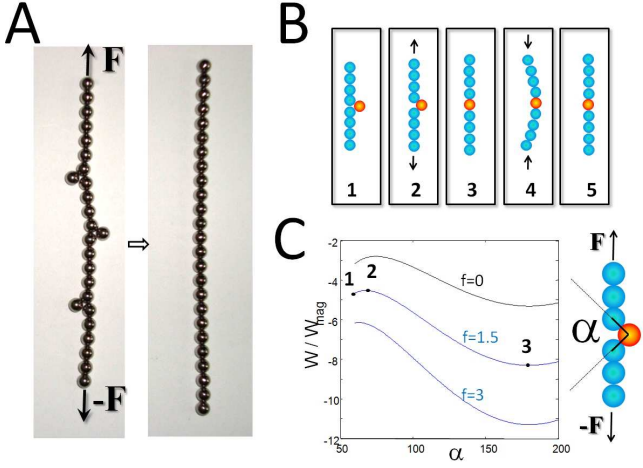}
\caption{ A) A macroscopic (Neodymium) magnet chain straightens
under a pulling force once a critical tension is reached. B) The
recombination process of a side-chain defect becoming absorbed
into the main chain during pulling (2,3). In the reverse cycle of
pushing the ends with the same force , the chain does not revert
but rather buckles on the large scale (3-5). C) The magnetic plus
potential energy of a long, stretched, magnetic chain under
various applied forces as function of the defect opening angle
$\alpha.$ At a critical force the boundary minimum at $\alpha=0$
becomes unstable and the chain relaxes in the straight chain ,
$\alpha=180{{}^{\circ}}$ global energy minimum. Due to the
asymmetry of the energy landscape the process is
irreversible and exhibits a configurational hysteresis. }%
\label{fig10}%
\end{figure}

As discussed in the main text, the lengthening of the aggregates can be
associated with the popping-in transition of defects under tension.

Here we consider a simple model system :\ a chain of magnetic
beads with a permanent magnetic moment $m$ and a single side
defect bead with radius $R$ (see Fig \ref{fig10}). The chain is
acted upon an external force $F$ between its ends that mimics the
tension when two beads in our "dimer-toy model", are sitting near
the potential maximum or compression when they are at the minima.

We assume a linear chain consisting of $2N+1$ beads along the y axis at
positions $2R\left(  0,\sin\dfrac{\alpha}{2}+k\right)  $ and $2R\left(
0,-\sin\dfrac{\alpha}{2}-l\right)  $ $,l,k=0,...N$ respectively and one singe
side chain bead at the middle position $2R\left(  \cos\dfrac{\alpha}%
{2},0\right)  $, and further that all the moments $m$ are all
pointing along the chain main axis (in the $y$ direction
${\mathbf{e}_{y}}=\left(  0,1\right)  )$ . Here
$\alpha$ stands for the angle at the defect. The latter is $\alpha=60%
{{}^\circ}%
$ if the 3 beads at the defect are in contact , i.e. the defect in
a "closed state"
and $\alpha>60%
{{}^\circ}%
$ when the defect starts to open up and straighten. At $\alpha=$ $180%
{{}^\circ}%
$ the defect fully immerses into the main chain and has essentially
disappeared. Therefore the angle $\alpha$ acts as a reaction coordinate
representing the state of the defect.

The potential plus dipole-dipole interaction energy of the system is up to a
$\alpha$-independent constant%

\begin{align*}
W  &  =2RF\sin\frac{\alpha}{2}-\frac{\mu_{0}m^{2}}{2\pi}2\sum_{k=0}^{N-1}%
\frac{3\left(  \mathbf{e}_{y}\cdot\mathbf{n}_{k}\right)  ^{2}-1}{r_{k}^{3}}\\
&  -\frac{\mu_{0}m^{2}}{2\pi}\sum_{l=0}^{N-1}\sum_{k=0}^{N-1}\frac{2}%
{r_{kl}^{3}}+const.
\end{align*}
with the unit bond vectors $\mathbf{n}_{k}=\frac{1}{\sqrt{k^{2}+2k\sin\frac
{1}{2}\alpha+1}}\left(  -\cos\dfrac{\alpha}{2},\sin\dfrac{\alpha}{2}+k\right)
$ between the main chain and the defect particle, $r_{k}$ the distance between
the defect bead and the $k-th$ main chain bead and $r_{kl}$ the distance
between the $k-th$ and $l-th$ bead in the top/bottom part of the main chain.
This can be rewritten in terms of $\alpha$ :%
\begin{align*}
\frac{W}{W_{mag}}  &  =\frac{2RF}{W_{mag}}\sin\dfrac{\alpha}{2}+2\sum
_{k=0}^{N-1}\frac{1-3\left(  \sin\dfrac{\alpha}{2}+k\right)  ^{2}/d\left(
\alpha,k\right)  }{d\left(  \alpha,k\right)  ^{3/2}}\\
&  -\sum_{l=0}^{N-1}\sum_{k=0}^{N-1}\frac{2}{\left(  2\sin\dfrac{\alpha}%
{2}+k+l\right)  ^{3}}%
\end{align*}

with $W_{mag}=\frac{\mu_{0}m^{2}}{4\pi R^{3}}$ and $d\left(  \alpha,k\right)
=k^{2}+2k\sin\frac{\alpha}{2}+1$ .

Note that even a chain with no tension applied can become unstable once a
critical distance between the center-beads is surpassed. For a very long
chain, $N\rightarrow\infty,$ the summation can be performed numerically and
the barrier is located at $\alpha=\alpha_{c}\approx75%
{{}^\circ}%
$ (for a 2 bead main chain at $\alpha_{c}\approx70%
{{}^\circ}%
$ ) . This corresponds to a surface-to-surface opening distance of
\[
\frac{D_{c}}{2R}=2\sin\left(  \frac{\alpha_{c}}{2}\right)  -1\approx0.22
\]
That is, for a main chain opening by a 22\% of the bead diameter the defect
becomes unstable.

For a simpler, analytically tractable case of only\ 3 bead recombination (2 in
the main-chain and one recombining) with $N=1$ and the energy simplifies to:%
\begin{align*}
\frac{W}{W_{mag}}  &  =fu-\frac{1}{4u^{3}}+2\left(  1-3u^{2}\right) \\
u  &  =\sin\dfrac{\alpha}{2}\text{ and }f=\frac{2RF}{W_{mag}}%
\end{align*}

$u$ is the half-distance between the main-chain beads. It goes from
$u_{1}=1/2$ for $\alpha=60%
{{}^\circ}%
$ and \ $u_{2}=\allowbreak1$ for $\alpha=180%
{{}^\circ}%
.$ It has extrema at%
\[
f+\frac{3}{4u^{4}}-12u=0
\]

For vanishing force $f=0$ we have the barrier at $u^{\ast}=$
$0.574\,35.$ For nonzero $f$, in the two limiting cases $u=u_1$
and $u=u_2$ the barrier disappears for $f_{1}=-6$ and $f_{2}=11.\,
\allowbreak25$. The negative force $f_{1}$ corresponds to the
critical tension for the side chain to "pop-in" and the positive
$f_{2}$ corresponds to the buckling compression for the bead to
"pop-out". The magnitude of the force necessary to create a defect
is almost twice as large as the force needed to absorb the defect
in the main chain.


\begin{thebibliography}{99}
\bibitem {NOTE2}Note, that the equation for $\Theta(t)$ is similar to the
Josephson equation for the dynamics of the superconducting phase difference in
a resistive current-driven $\pi-$junction circuit \cite{Bulaevskii}. The above
quantities $\dot{\Theta}$, $\omega$ and $\omega_{c}\sin\Theta$ play the role
of voltage (on the resistive element), driven external current and Josephson
current, respectively. The equation appears also in other systems involving
particle coupling to rotating fields and this principle has been used for
micro-sensing by optical torque driven beads \cite{ExcitableParticles}.
\end{thebibliography}
\end{document}